\input psfig
\documentstyle[twoside,fleqn,espcrc2]{article}

\newcommand{\AmS}{{\protect\the\textfont2
  A\kern-.1667em\lower.5ex\hbox{M}\kern-.125emS}}

\title{Evidence for quenched chiral logs\thanks{Talk presented by H. Thacker.
Supported by DOE grant DE-FG02-97ER41027}}

\author{W. Bardeen,\address{Fermilab, P.O. Box 500, Batavia, IL 60510}%
        A. Duncan,\address{Dept. of Physics and Astronomy,University of Pittsburgh, Pittsburgh, PA 15260}%
        E.~Eichten,$^{\rm a}$
        and 
        H.~Thacker\address{Dept. of Physics, University of Virginia, Charlottesville, VA 22901}}

\begin{document}

\begin{abstract}
 
Using the pole shifting procedure of the modified quenched 
approximation (MQA) to cure
the exceptional configuration problem, accurate hadron spectrum 
calculations can be obtained at very light quark mass. Here we use
the MQA to extend and improve our previous investigation of chiral
logs in the pion mass. At $\beta = 5.7$ for Wilson fermions, we
see clear evidence for quenched chiral logarithms in the pion
mass as a function of quark mass. The size of the observed chiral
log exponent $\delta$ is in good agreement with the value obtained from a
direct calculation of the $\eta'$ hairpin diagram.

\end{abstract}

\maketitle

\section{INTRODUCTION}

Some time ago Sharpe and Bernard and Goltermann\cite{Sharpe}
showed that the quark mass dependence of certain 
physical quantities near the chiral limit should serve as a
particularly incisive test of the difference between quenched
and full QCD. In particular, the behavior of the pion
mass as a function of the quark mass in quenched QCD is expected
to be non-analytic at $m_q=0$, with an anomalous exponent $\delta$,
\begin{equation}
\label{eq:chlog}
m_{\pi}^2 \propto (m_q)^{\frac{1}{1+\delta}}
\end{equation}
The value of the chiral log exponent in the pion mass should 
be related to the coefficient of the $\eta'$ hairpin insertion by
\begin{equation}
\label{eq:delta}
\delta = \frac{m_0^2}{48\pi^2f_{\pi}^2}
\end{equation}
Calculations we have reported previously\cite{Lat96} showed no evidence
for chiral logs in the pion mass at $\beta=5.7$ over the range of hopping
parameters used in that calculation, from .1610 to .1680
(pion masses of .650 to .254 in lattice units). The value
of the exponent $\delta = .015(47)$ was obtained, i.e.
consistent with zero. The results reported here are 
completely consistent with the previous negative results. We
find that the value of $\delta$ obtained in this region
from the hairpin coefficient varies from $\delta<.01$ at 
$\kappa=.161$ to $\delta = .045(5)$ at $\kappa=.168$.
Even when extrapolated to the chiral limit, the exponent
is only $\delta=.053(7)$. 

In this present calculation,
we have used the pole-shifting ansatz of the modified
quenched approximation (MQA) \cite{MQA} to resolve the
exceptional configuration problem. This allows us to 
greatly improve our statistics at light quark masses
and to extend the calculations much closer to critical
hopping parameter.
The pion mass at these lighter quark mass values is much more
sensitive to the chiral logs and our errors are small
enough to see a significant effect even though $\delta\approx .05$.
At the lightest three quark masses (.1683, .1685, and .1687)
we find clear evidence for a deviation from linearity
at just the level and functional form 
expected from the hairpin calculation. The agreement between the hairpin
result and the pion mass behavior provides significant numerical
support for the theoretical picture of chiral
logarithms and their connection to $\eta'$ loops in the quenched
approximation.

Both the overall smallness of the chiral log parameter 
and the strong quark mass dependence are somewhat unexpected,
in light of the fact that (a) the physical $\eta'$ mass gives
$\delta\approx.17$, and (b) in the chiral limit,
the hairpin mass insertion is 
related by the Witten-Veneziano formula to the topological
susceptibility of pure glue (without quarks) and should
therefore be rather insensitive to the quark mass (at
least for quark masses much smaller than the QCD scale).
In an effort to understand the origin of this suppression 
and quark mass dependence, we carried out another calculation of
the hairpin diagram using a clover improved quark action.
(The clover coeffient used here is $C_{sw}=1.57$.) The two
main effects of clover improvement are (1) to increase the
hairpin coefficient $m_0^2$ by more than a factor of 2,
and (2) to decrease the slope of $m_0$ as a function of
quark mass by about a factor of 2 (where quark mass for
the clover case is rescaled so that equal quark mass 
corresponds to equal pion mass). This strongly suggests
that both the suppression and the strong quark mass dependence
of $m_0$ are lattice artifacts.
Thus we would expect the chiral log parameter $\delta$ 
to be quite sensitive
to lattice spacing and to increase substantially at higher
$\beta$ values. Recent results on the pion mass from 
the $CP-PACS$ collaboration\cite{CP}
indicating that $\delta\approx .1$ at $\beta=6.47$ are quite consistent
with this conclusion.\cite{future}

\begin{table}
\caption{$\eta'$ mass $m_0$, predicted chiral log parameter $\delta$,
and pion mass.}
\label{tab:deltalimits}
\begin{tabular}{|c|c|c|c|}
\hline
$\kappa$& $m_0$ & $\delta$ & $m_{\pi}$ \\
\hline
$.1687$ & $        $ & $       $ & $.165(5)$ \\
$.1685$ & $.383(22)$ & $.050(6)$ & $.195(4)$ \\
$.1683$ & $        $ & $       $ & $.221(3)$ \\
$.1680$ & $.371(20)$ & $.045(5)$ & $.254(2)$\\
$.1675$ & $.346(16)$ & $.038(3)$ & $.298(2)$\\
$.1667$ & $.316(14)$ & $.030(3)$ & $.356(2)$\\
$.1650$ & $.268(13)$ & $.019(2)$ & $.458(1)$\\
$.1630$ & $.221(13)$ & $.011(1)$ & $.558(1)$\\
\hline
\end{tabular}
\end{table}

In Table I the results for $m_0$ are presented. Also listed
are the resulting values of the chiral log exponent $\delta$,
using Eq. (\ref{eq:delta}), with $f_{\pi}$ calculated on the same
ensemble. In the last column, we list the
values of the pion mass obtained from the valence pion propagator.

As a function of $\kappa^{-1}$, the values of $m_{\pi}^2$ obtained
from Table I are quite linear for the heavier masses, but show
a significant deviation from linearity for $\kappa > .1680$.
This deviation has both the form and the magnitude predicted
by the chiral log formula (\ref{eq:chlog}) combined with the
hairpin values for $\delta$. A direct fit of $m_{\pi}^2$
to the formula (\ref{eq:chlog}), including values from $\kappa
=.165$ to $\kappa=.1687$ gives $\delta=.058(39)$, i.e.
significantly different from zero and consistent with the
hairpin results. However, this direct fitting procedure with constant
$\delta$ has
several drawbacks. The rather large error reflects the observed fact
that the value of $\delta$ changes significantly over the fit
range. Also, the value of $\kappa_c$ must be taken as one of
the fitting parameters, and there is a strong correlation
between $\kappa_c$ and $\delta$. 

A more accurate way to observe the presence and magnitude of the
chiral logs is to consider the ratio of differences:
\begin{equation}
\label{eq:ratio}
\frac{\Delta\left(m_{\pi}^{2(1+\delta)}\right)}{\Delta\left(
\kappa^{-1}\right)}
\end{equation}
where the differences are computed between adjacent values of
$\kappa$. With the correct choice of $\delta$, this ratio 
should be constant. Note that calculating this ratio does not require a
determination of $\kappa_c$. Another advantage is that the
values of $m_{\pi}$ at different $\kappa$'s are positively
correlated, so that errors on the difference are typically
30-50\% smaller than on the (mass)$^2$ itself. The errors
are computed by a single elimination jackknife. To see the
chiral log effect, we first plot in Fig.1 the ratio 
(\ref{eq:ratio}) for $\delta = 0$. The horizontal dashed line
represents the result of the best linear fit to $m_{\pi}^2$
for the $\kappa$ range $.163$ to $.168$. The three points
at the lightest quark mass, $<25$ MeV, are obtained from the
three differences computed between the four kappa values
.1680, .1683, .1685, and .1687. These points show a
significant chiral log effect which is consistent with
the hairpin expectations. To see this we plot in Fig. 2
the ratio (\ref{eq:ratio}) with values of $\delta$
taken from interpolating the hairpin results in Table I. With this 
choice of $\delta$, the ratio's are consistent with 
a constant. To see how sensitive this plot is to the
chosen values of $\delta$, we plot in Fig. 3 the ratios
(\ref{eq:ratio}) with a mass independent value of
$\delta=.1$. The behavior of the ratio is now completely
inconsistent with a constant, illustrating that a
value as large as $\delta=.1$ is ruled out by the data.

\begin{figure}
\vspace*{3.6cm}
\includegraphics{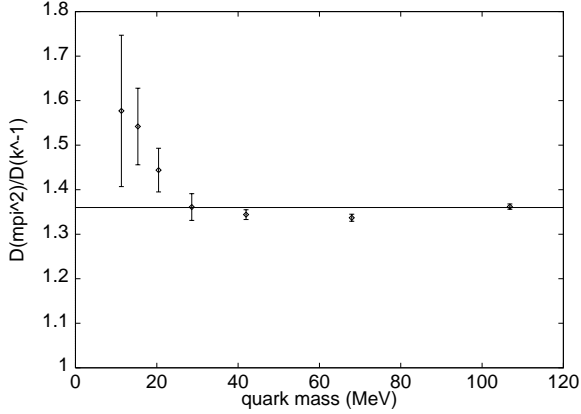}
\vspace{0.5cm}
\caption[]{The ratio $\Delta\left(m_{\pi}^2\right)/\Delta\left(
\kappa^{-1}\right)$, i.e. Eq. \ref{eq:ratio} with $\delta=0$.}
\end{figure}

\begin{figure}
\vspace*{3.6cm}
\includegraphics{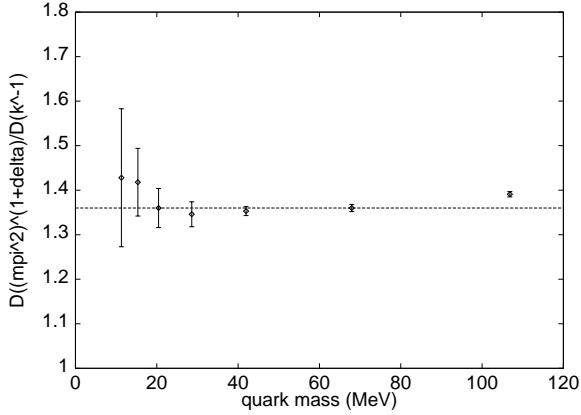}
\vspace{0.5cm}
\caption[]{The ratio $\Delta\left(m_{\pi}^{2(1+\delta)}\right)/\Delta\left(
\kappa^{-1}\right)$ with $\delta$ from Table I.}
\end{figure}

\begin{figure}
\vspace*{4.0cm}
\includegraphics{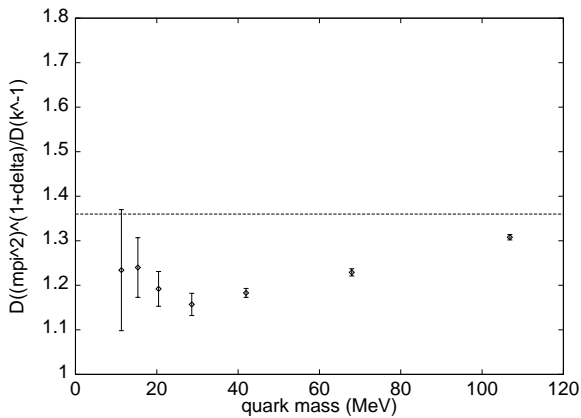}
\vspace{0.5cm}
\caption[]{The ratio $\Delta\left(m_{\pi}^{2(1+\delta)}\right)/\Delta\left(
\kappa^{-1}\right)$ with $\delta=0.1$.}
\end{figure}

Finally, we present in Fig. 4 the values of $m_0$ (in
lattice units) obtained from the clover improved action,
compared with those obtained from the unimproved Wilson
action.
As discussed above, the clover improved value is 
substantially larger in magnitude and the quark mass
dependence is significantly reduced.\cite{future}

\begin{figure}
\vspace*{4.0cm}
\includegraphics{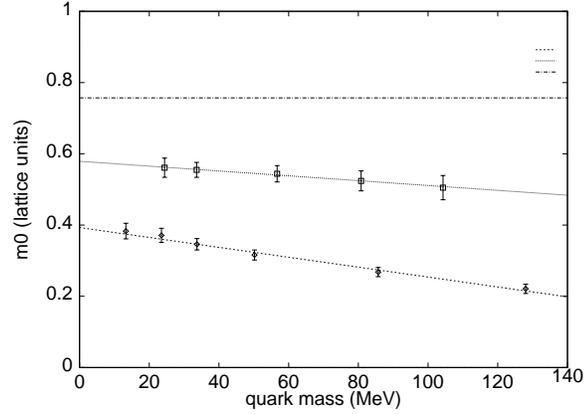}
\vspace{0.5cm}
\caption[]{Comparison of $m_0$ for Wilson action (lower line) and
clover action with $C_{sw}=1.57$ (middle line). The upper line is the result of a linear
extrapolation to zero slope.}
\label{fig:hp_raw}
\end{figure}

\end{document}